\documentclass[conference]{IEEEtran}
\IEEEoverridecommandlockouts
\usepackage{algorithmic}
\usepackage{booktabs}
\usepackage{setspace}
\usepackage{float}
\usepackage[utf8]{inputenc}
\usepackage{multirow}
\usepackage[table,xcdraw]{xcolor}
\usepackage[sorting=none]{biblatex}

\usepackage{amsmath,amssymb,amsfonts}
\usepackage{fdsymbol}
\usepackage{graphicx}
\usepackage{subcaption}
\usepackage{adjustbox}

\usepackage{caption}
\usepackage{threeparttable}
\usepackage{xcolor}
\usepackage{soul}
\usepackage{textcomp}
\usepackage{fancyhdr}
\def\BibTeX{{\rm B\kern-.05em{\sc i\kern-.025em b}\kern-.08em
    T\kern-.1667em\lower.7ex\hbox{E}\kern-.125emX}}

\fancypagestyle{FirstPage}{
\lfoot{979-8-3503-3748-8/23/\$31.00 \textcopyright{}2023 IEEE} 
}

\begin{document}

\title{Robust CNN-based Respiration Rate Estimation for Smartwatch PPG and IMU\\
}

\author{Kianoosh Kazemi$^1$,
Iman Azimi$^2$,
Amir M. Rahmani$^{2,3}$,
and Pasi Liljeberg$^1$\\
\textit{$^1$Department of Computing, University of Turku, Turku, Finland}\\
\textit{$^2$Department of Computer Science, University of California, Irvine, USA}\\
\textit{$^3$School of Nursing, University of California, Irvine, USA}\\
{\{kianoosh.k.kazemi,  pasi.liljeberg\}@utu.fi, \{azimii, a.rahmani\}@uci.edu}}

\maketitle
\begin{abstract}
Respiratory rate (RR) serves as an indicator of various medical conditions, such as cardiovascular diseases and sleep disorders.
Several studies have employed signal processing and machine learning techniques to extract RR from biosignals, such as photoplethysmogram (PPG). These RR estimation methods were mostly designed for finger-based PPG collected from subjects in stationary situations (e.g., in hospitals). In contrast to finger-based PPG signals, wrist-based PPG are more susceptible to noise, particularly in their low frequency range, which includes respiratory information. Therefore, the existing methods struggle to accurately extract RR when PPG data are collected from wrist area under free-living conditions. The increasing popularity of smartwatches, equipped with various sensors including PPG, has prompted the need for a robust RR estimation method. In this paper, we propose a convolutional neural network-based approach to extract RR from PPG, accelerometer, and gyroscope signals captured via smartwatches. Our method, including a dilated residual inception module and 1D convolutions, extract the temporal information from the signals, enabling RR estimation. Our method is trained and tested using data collected from 36 subjects under free-living conditions for one day using Samsung Gear Sport watches. For evaluation, we compare the proposed method with four state-of-the-art RR estimation methods. The RR estimates are compared with RR references obtained from a chest-band device. The results show that our method outperforms the existing methods with the Mean-Absolute-Error and Root-Mean-Square-Error of 1.85 and 2.34, while the best results obtained by the other methods are 2.41 and 3.29, respectively. Moreover, compared to the other methods, the absolute error distribution of our method was narrow (with the lowest median), indicating a higher level of agreement between the estimated and reference RR values. 

\end{abstract}


\section{Introduction} \label{introduction}

Respiratory rate (RR), commonly known as the breathing rate, denotes the quantity of breaths taken by an individual in a minute. Typically, grown-ups' average resting respiratory rate falls between 12 and 20 breaths per minute \cite{clevelandclinic}. Deviations from the standard respiratory rate can indicate a person's physiological condition, such as anxiety, hypoxia, hypercapnia, and respiratory acidosis \cite{rolfe2019importance}. Many studies have indicated that respiration rate is a crucial indicator of cardiac arrest and patient deterioration \cite{cooper2014respiratory, cretikos2008respiratory}. Therefore, monitoring RR is crucial to assessing patients' health status in hospitals, clinics, and homes. 




In recent years, several studies have been proposed to estimate respiratory rate using physiological signals such as Photoplethysmogram (PPG), Electrocardiogram (ECG), and accelerometer (ACC) signals \cite{aqajari2021end, bian2020respiratory, charlton2017breathing, huang2021novel, rathore2022mrnet}. The existing approaches to extract respiration rate from biosignals can be categorized into three main groups as Waveform Analysis, Deep Learning approaches, and Conventional Machine Learning. Waveform Analysis methods include several steps, such as filtering, signal quality evaluation, time/frequency domain analysis, respiratory-induced waveforms extraction, features extraction based on fiducial points, and feature fusion. Nicholas et al. \cite{huang2021novel} proposed an RR estimation approach based on the fusion of PPG and ACC-derived respiratory rates. From PPG signals, they extracted surrogate respiration waveforms from distinct modulations \cite{meredith2012photoplethysmographic} (i.e., amplitude (AM), intensity (BW), and frequency (FM) modulation ). Simultaneously, they used time-frequency spectrum extraction, followed by spectral peak-tracking for ACC-based RR estimation. In the final step, PPG- and ACC-based RR estimates were fused based on their quality. In another study, an RR estimation method based on ACC and Gyroscopes (GYR) was presented \cite{hernandez2015biowatch}. The method included several processing steps, such as accelerometer components normalization, filtering techniques to isolate relevant information, sensor components aggregation, and frequency-domain-based RR estimation. These methodologies are low-cost and easy to implement, but they require parameters to be tuned manually, optimized, and handcrafted rules customized for specific patient populations causing a lack of robustness to the presence of noise \cite{aqajari2021end, bian2020respiratory}.


Conventional Machine Learning methods were also employed for RR estimation. These studies proposed machine learning methods fed by time and frequency features extracted from PPG signals. For example, Kapil et al. \cite{rathore2022mrnet} derived frequency domains (such as Fourier spectrum, power spectral density, and Spectral entropy) and morphological features (e.g., autocorrelation, autoregression coefficient, Hjorth complexity, and peak-to-valley amplitude) from PPG and different modalities. Then, they developed machine learning models, including Random Forest, Gradient Boosting, and Support Vector Machine, to predict the RR. 

In addition, deep learning-based approaches have been developed for RR estimation. These methods efficiently extracted temporal and spatial information from 1-D time series and estimated RR automatically from raw biosignals (i.e., PPG) \cite{aqajari2021end, bian2020respiratory}. Aqajari et al.  \cite {aqajari2021end} employed cycle generative adversarial networks (GAN) to construct respiration signals. They leveraged the ability of cycle GAN to translate PPG signals to respiration signals and incorporated the error between the reconstructed and ground truth signals into an optimization function to ensure the accuracy of the reconstructed signal for RR estimation. In another work \cite{bian2020respiratory}, a deep learning approach was introduced for automatic and accurate RR estimation. They proposed an end-to-end learning model based on the ResNet block as a backbone attached to fully-connected layers to estimate RR from raw PPG signals.   

The existing machine learning and deep learning methods were developed to extract RR from PPG signals. However, they mostly fail when the signal-to-noise ratio of PPG is low and subsequently lead to incorrect RR extraction. These methods were designed and tested on PPG signals captured in a stationary position (i.e., minimum or no motion) via, for example, pulse oximeter worn by a patient in an intensive care unit \cite{karlen2013multiparameter, pimentel2016toward}.
In addition, the state-of-the-art RR extraction methods were developed for fingertips-based PPG signals. It is worth noting that the choice of measurement site significantly impacts the quality of the collected PPG signal. Wrist-based PPG signals are more susceptible to noise and motion artifacts when compared to their finger-based counterparts. The difference is particularly significant in low-frequency range (i.e., RR frequency range) \cite{hartmann2019quantitative}. Consequently, RR estimation using wrist-based PPG signals is relatively more challenging than the estimation conducted using finger-based PPG signals \cite{longmore2019comparison}.

Recently, there has been an emerging trend in the field of wearable technology, with a specific focus on smartwatches and fitness trackers. These devices are equipped with various sensors – including PPG and Inertial Measurement Units – enabling efficient and ubiquitous acquisition of various physiological signals from the wrist area. Studies showed that these signals carry useful information related to respiration-induced motion \cite{sun2017sleepmonitor, lee2020respiration}, although they are susceptible to noise due to their collection from the wrist and potential hand movements during the user's daily routines in free living conditions. We believe that an RR estimation method is necessary for wrist-based devices to integrate these signals, leverage their temporal and spatial information, and deliver a robust RR.


In this paper, we propose a deep learning-based approach to extract RR from PPG, ACC, and GYR signals collected from smartwatches in free-living conditions. The learning model includes a dilated residual inception module and 1D convolutions. We exploit Independent Component Analysis (ICA) method to obtain respiratory signals from ACC and GYR. Then, the PPG and the extracted respiratory signals are fed to the network to capture the temporal information. Our method is evaluated using PPG, ACC, and GYR signals collected from 36 subjects under free-living conditions for 24 hours using Samsung Gear Sport watches. The performance of the proposed method is evaluated and compared with state-of-the-art RR estimation methods.
In summary, our contributions are as follows:
\begin{itemize}
    \item Proposing multi-scale convolution incorporated with residual inception networks for RR extraction from wrist-based physiological signals.
    \item Evaluating the proposed method using smartwatches data collected in free-living conditions.
    \item Assessing the proposed method's performance in terms of MAE, RMSE, Parameter Count, Mean bias, and confidence interval compared to the state-of-the-art methods.
\end{itemize}
Section \ref{Dataset} describes the dataset used in this study for the methods' evaluation. Section \ref{Pipeline} outlines the proposed RR estimation method. In Section \ref{discussion}, we present the setup and experimental results. Finally, the paper concludes in Section \ref{conclusion}.

\section{Dataset}\label{Dataset}
\begin{figure}[htbp]
\includegraphics[height=5.5cm ,width=7.0cm]{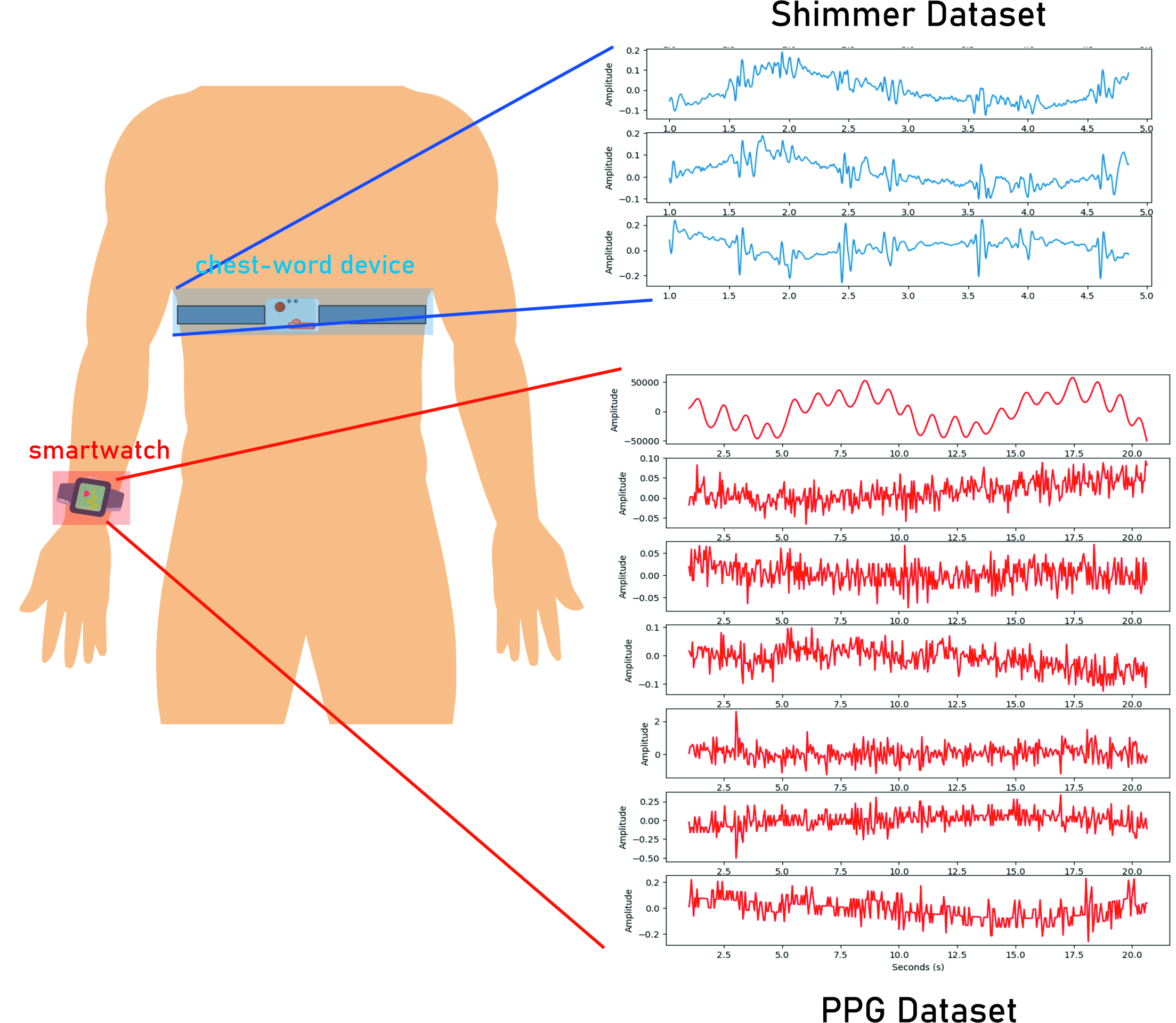}
\caption{The monitoring system employed for dataset collection consisting Samsung Gear Sport smartwatches and Shimmer3 ECG devices.}
\label{fig:1}
\end{figure}

This work utilized a PPG, ACC, and GYR dataset collected as a part of a health monitoring study \cite{mehrabadi2020sleep}. In the study, the participants were instructed to wear Samsung Gear Sport smartwatches on their non-dominant wrists and a Shimmer3 device on the chest (Fig.~\ref{fig:1}). The data collection was conducted continuously for 24 hours under free-living conditions while the participants engaged in their typical daily activities.

The recruitment and data collection were conducted in southern Finland throughout July-August 2019. The recruitment process was initiated among the University of Turku's students, and additional participants were recruited using a snowball sampling technique, ultimately resulting in 36 participants. All participants were healthy individuals, with equal representation of both males and females. Exclusion criteria were applied during the recruitment process to ensure a homogeneous sample. The exclusion criteria included restrictions on the use of wearable devices while working, physical activity limitations, having cardiovascular disease, and the presence of symptoms of illness at the time of recruitment. 

The Shimmer3 device \cite{shimmer} was used to collect chest Accelerometer measurements, used in our analysis as the reference signal (ground Truth) for RR extraction. Chest-worn accelerometers are proven to be a robust and noise-resilient method for RR estimation \cite{schipper2021estimation, doheny2020estimation}. The Shimmer3  device is a lightweight and compact apparatus programmed to record Tri-axial accelerometer data continuously \cite{shimmer}. The device is equipped with adequate internal storage and battery life to enable continuous data recording for 24 hours. The Shimmer device was configured to capture data at a sampling frequency of 512 Hz \cite{peltola2012role}.

The Samsung Gear Sport smartwatch \cite{samsung} was employed to acquire wrist PPG and triaxial gyroscope/accelerometer signals. Samsung Gear Sport Watch is compact and lightweight, weighing 67 g with the strap. A waterproof smartwatch powered by the open-source Tizen operating system with a three-day battery life. It is equipped with PPG and inertial measurement unit (IMU) sensors that collect data at a frequency of 20 Hz.

During the study, the Declaration of Helsinki and the Finnish Medical Research Act (No 488/1999) were followed. University of Turku for Human Sciences Ethics Committee has approved this study (No 44/2019). Prior to giving their written consent, participants were informed orally and in writing. There was no obligation for participants to participate in the study, and they could withdraw from it if they did not wish to continue taking part in the study.

\section{Deep Learning-Based RR Estimation Pipeline}\label{Pipeline}
\begin{figure}[htbp]
\includegraphics[height=2.35cm ,width=8.75cm]{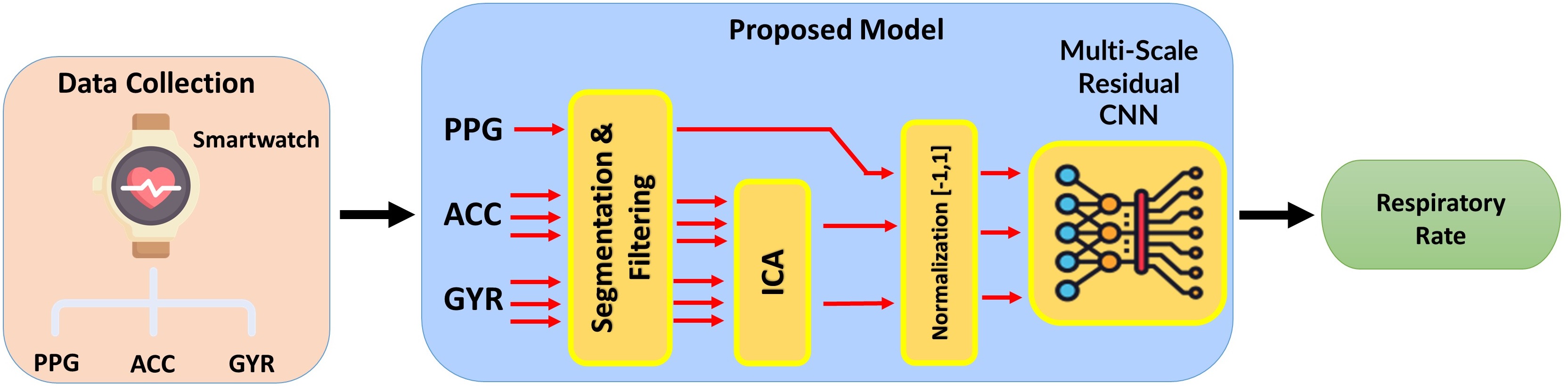}
\caption{Proposed RR estimation method}
\label{fig:2}
\end{figure}

In this section, we introduce our deep learning-based RR estimation method developed for wrist-band devices, equipped with PPG, accelerometer, and gyroscope sensors. The proposed data analysis pipeline (shown in Fig. \ref{fig:2}) receives PPG, ACC, and GYR signals, as inputs, and estimates the fused RR values, as outputs. The method pipeline consists of different components to first segment the input data and discard distorted signals. Second, it extracts respiration signals from ACC and GYR. Finally, the Multi-Scale Residual CNN extracts automatic features from the signals to estimate RR values. In the following, we describe the major components in detail.
\subsection{Segmentation and Filtering}
\subsubsection{Segmentation:} We first segment the input PPG, ACC, and GYR signals. The collected PPG, ACC, and GYR signals are nonstationary in terms of respiration rate and noise level, meaning that respiration rate and noise levels differ while monitoring over time. Therefore, the signals are divided into (quasi) stationary segments, assuming a fixed respiration rate and noise level within the segment. The length of the segments should be sufficient to enable accurate waveform analysis but not too long to lose the (quasi) stationary nature. According to the literature, the most suitable duration for estimating the respiratory rate and respiration signals can vary between 30 seconds to 90 seconds \cite{schafer2008estimation}. Longer window sizes tend to result in lower errors \cite{karlen2013multiparameter}, whereas shorter windows are more computationally efficient and stable for RR estimation. In our analysis, 32-second segments are selected. Therefore, 32-second windows of raw PPG, triaxial ACC, and GYR signals are randomly selected. 
\subsubsection{Filtering:} This step aims to eliminate the segments of the PPG, ACC, and GYR signals obtained during users' activities. Wrist-based signals (i.e., PPG, ACC, and GYR) also capture motion artifacts and hand movement, masking the signals' oscillation due to RR. The frequency components of these noises often overlap with the breathing frequency range. To address this issue, a quality assessment technique was employed, consisting of five morphological and frequency domain characteristics (i.e., power spectral density, Interquartile range,  heart cycle energy, correlation and Euclidean distances between cardiac cycles and template) \cite{feli2023energy} and a one-class support vector machine method, to effectively filter out the motion-corrupted portions of the PPG, ACC, and GYR signals.

\subsection{ICA and Normalization:} We employ an ICA method to extract the first principal axes of ACC and GYR. The ICA technique allows the separation of mixed signals into their underlying independent components \cite{kim2006motion}. The accelerometer and gyroscope sensors recorded multiple types of signals, such as breathing data, motion, and environmental noises \cite{lee2020respiration}. These signals are assumed to have independent distribution. In this regard, ICA was used to extract two respiration-related signals from raw ACC and GYR signals. Then, the  PPG  and the extracted respiratory signal from ACC and GYR signals are scaled between -1 and 1 to be used for training and testing our machine learning model.


\subsection{Multi-Scale Residual CNN}
\begin{figure}[htbp]
\includegraphics[height=5.25cm ,width=8.85cm]{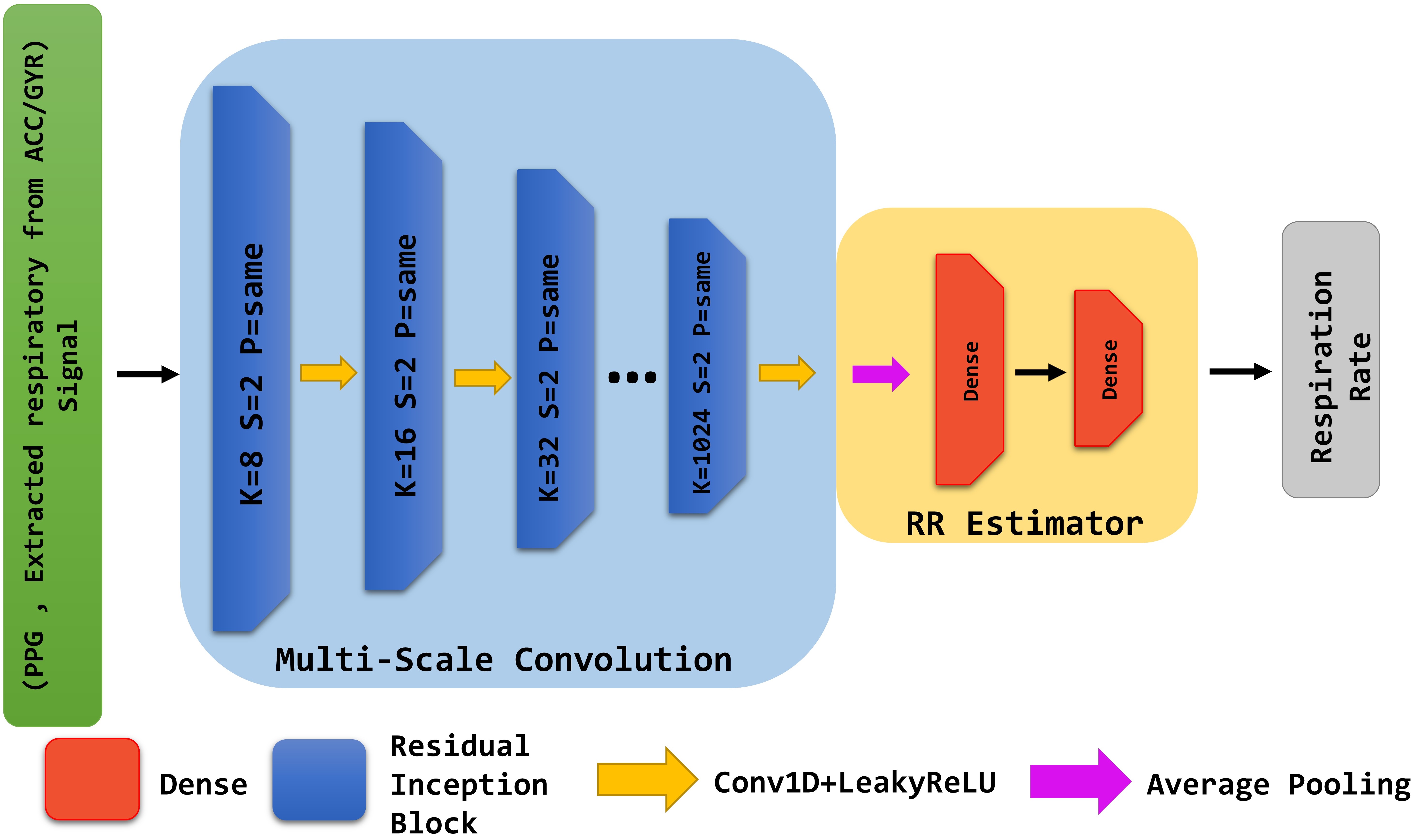}
\caption{Deep neural network architecture}
\label{fig:3}
\end{figure}

We propose a deep neural network consisting of two distinct modules, a multi-scale convolution and respiratory rate estimator, to estimate the RR. A view of our model architecture is depicted in Fig.~\ref{fig:3}. We briefly outline the two modules in the following.

\textbf{1) Multi-Scale Convolution:} We develop a CNN-based layer to receive the three signals and extract automatic features. By inspiring from the multi-scale convolution method implemented in the grid modules of various inception networks \cite{szegedy2017inception}, we incorporate a dilated residual inception module in our proposed model to effectively capture signal features at different resolutions and reduce the problem of vanishing gradient significantly \cite{he2016deep}. In addition, dilated residual inception blocks provide larger receptive fields without significantly increasing parameters. Our approach involves three parallel branches, where the input signals go through three separate channel-wise convolutional layers before being concatenated at the end. Each convolutional layer utilizes diverse filter sizes, enabling the acquisition of appropriate weights for each convolutional resolution, thereby determining the frequency of recurring features present in the input signals. This information is then combined for subsequent modules. Integration of the dilated residual inception block leads to more robust feature extraction, ultimately leading to a smaller architecture.

This module is augmented by incorporating a Batch Normalization, a Leaky ReLU activation layer with a slope of 0.2, and 1D convolution layers. Each layer of the 1D convolution has a kernel size of three and a stride of two. Initially, the convolution layer has a filter size of 8, which is gradually increased by a factor of two until it reaches 1024. In order to enhance the efficiency of the training process, the utilization of strided convolution is preferred over max-pooling for downsampling  \cite{springenberg2014striving}. The downsampling procedure results in a reduction of the input size while amplifying the count of filters for every layer by a factor of two until the number of filters reaches 1024 for the subsequent levels.

The Multi-Scale Convolution is fed with the raw PPG and two respiratory signals extracted from the ACC and GYR signals using the ICA method. Our model inputs can be expressed as: $X$ = $(x^{(11)}$, $x^{(12)}$, $x^{(13)}$, $RR_{r}^{(11)})$, $(x^{(21)}$, $x^{(22)}$, $x^{(23)}$, $RR_{r}^{(21)})$, $...$ , $(x^{(n1)}$, $x^{(n2)}$, $x^{(n3)}$, $RR_{r}^{(n1)})$, where $(x^{(i)}$, $RR_{r}^{(i)}$ $\in$ $R^n)$. The Multi-Scale Convolution downsamples the input signals  $x(i)$ to obtain a compressed feature vector $Z^{(i)}$. The compressed feature vector is represented by \eqref{eq1}:
\begin{equation}
    Z^{(i)} = B_1\left( x^{(i)};\theta_1\right) \label{eq1}
\end{equation}
Where the $\theta_1$ is the parameters in the multi-Scale convolution denoted as B1.

\textbf{2) Respiratory Rate Estimator:} We leverage a fully connected layer to estimate RR values. The output of the convolution layers is fed to a global average pooling layer converting the downsampled extracted features from the residual blocks to a single dimension. Then, two sets of activation layers and a fully-connected layer shrink the input to sizes of 64 and 1, respectively, denoting as estimated RR. The dense layers can be expressed as:
\begin{equation}
    RR_{e}^{(i1)} = B_2\left( z^{(i)};\theta_2\right)\label{eq2}
\end{equation}
where the $\theta_2$ is the parameters in the dense layer denoted as  B2.
The optimization process for the architecture minimizes the $SmoothL$ loss between the estimated values $RR_{e}^{(ni)}$ and the reference values $RR_{r}^{(ni)}$. The loss function is defined as follows:
\begin{equation}
    L(X) = \sum_{i=1}^m SmoothL(RR_{diff})
\end{equation}
\begin{equation}
    y_{diff} = RR_{r}^{(i1)}-RR_{e}^{(i1)}
\end{equation}
\begin{equation}
    SmoothL_1(RR_{diff}) = \begin{cases}
  0.5(RR^2_{diff}) & abs(RR_{diff})<1\\    
  abs(RR_{diff})-0.5 & \text{otherwise }    
\end{cases}
\end{equation}
\section{Experimental  Results}\label{discussion}
\subsection{Model Training and Testing}
The proposed method is evaluated using raw PPG, ACC, and GYR signals of 36 healthy individuals collected for 24 hours under free-living conditions. An inter-patient test is conducted by selecting training and testing data from separate participants. We perform this test to ensure the model's generalization and prevent data leakage between train and test datasets. The training phase employs the PPG, ACC, and GYR data of 26 participants, comprising 5,300,000 32-second segments.
A total of 140,000 32-second segments from the remaining 10 participants are used in the testing phase. The watches were programmed to collect PPG data for 24 h at the sampling frequency of 20 Hz. In addition, Shimmer 3 collected data at a sampling rate of 512. We upsampled the PPG signals to 100 Hz and downsampled the Simmer 3 data to 100 Hz to unify the sampling frequencies. The upsampling and downsampling processes employed a conventional linear interpolation technique, a method commonly utilized for signal frequency conversion. In this method, a line is fitted between each pair of data points. Then, based on the upsampling or downsampling rate, new data points are fitted on the line. The proposed method is evaluated by comparing the predicted labeling values with the true labeling values. We compare our method with the state-of-the-art RR estimation methods, trained and tested with the same datasets.

The model training was carried out on a Linux machine powered by an AMD Ryzen Threadripper 2920X 12-Core processor, an NVIDIA TITAN RTX GPU, and 126GB of RAM. To develop the model, we used Tensorflow (version 2) and Keras API on Python. During the training experiment, the model was trained for 100 epochs, where the number of steps per epoch was set at 60, and an early stopping technique was employed to prevent over-fitting. Moreover, the Cosine Decay learning rate scheduler was adopted to yield the best results.

\subsection{Ground truth Respiration Rate (RR) }\label{section_RR_estimation}
We obtain the ground truth RRs values from ACC signals collected from a chest-band device (i.e., Shimmer3). The ground truth values are used as labels for model training and testing. 
The respiration rates are estimated from triaxial ACC signals using the method presented by Sun et al. \cite{sun2017sleepmonitor}. The method included 1) a preprocessing method to remove noise from the accelerometer data while preserving respiration-related fluctuations, 2) a Fast Fourier Transform method to obtain RR frequency, and 3) a multi-axis fusion approach to improve the estimates. The fusion approach consisted of a Kalman filter to integrate RR estimates from different X, Y, and Z axes, considering historical information. 

\subsection{Base-line methods}
We compare the proposed method with four existing RR estimation methods. We briefly describe these methods in the following.

\begin{table*}[!htp]
\caption{The summary of MAE, RMSE Performances, the 95\% confidence interval and mean difference for all methods}
\begin{center}
\begin{tabular}{ c c c c}
\toprule
\textbf{Method }                        & \textbf{MAE}                   & \textbf{RMSE}                   &  \textbf{Mean Bias}                 \\ \midrule
Proposed                            &    \textbf{1.85  $\pm$ 0.40 }                &     \textbf{2.34  $\pm$ 0.30 }              &\textbf{ -0.63}               \\ 
SF                                  &   2.91  $\pm$ 0.60                  &        3.52   $\pm$ 0.50            &     2.3                  \\ 
CNN                                 &  2.41 $\pm$ 0.60                    &    3.29 $\pm$ 1.20                  &                    1.02              \\ 
CycleGAN                            &    2.84  $\pm$ 0.90                 &      3.63 $\pm$ 0.80                &                       3.7              \\ 
\multicolumn{1}{c}{A-P Synthesis} & \multicolumn{1}{c}{2.90 $\pm$ 0.70} & \multicolumn{1}{l}{  3.96 $\pm$ 0.60} & \multicolumn{1}{c}{-2.2 } \\ \bottomrule
\end{tabular}
\label{tab1}
\end{center}
\end{table*}

\textbf{Smart Fusion (SF)}: A machine learning is used in \cite{rathore2022mrnet} to estimate RR from PPG signals. The method first extracts respiratory signal modalities (i.e., AM, FM, and BW) from PPG signals. Then, frequency-based features (e.g., Fourier spectrum, power spectral density, and Spectral entropy) are derived from individual respiratory signals. In addition to the frequency-based features, morphology-based features such as autocorrelation, autoregression coefficient, Hjorth complexity, and peak-to-valley amplitude features are extracted to improve the method's performance. Three different machine learning models (i.e., support vector machines, Random Forest, and Gradient Boosting) are developed to fuse these extracted features, and their predicted absolute errors are used to generate weights for each modality. These weights are calculated based on an expression that takes into account the predicted absolute error and an empirically determined constant $\tau$, which is set at 5.

\textbf{CNN}: The CNN-based model proposed in \cite{bian2020respiratory} is used to predict the respiration rate from raw PPG signals. The model's architecture consists of ResNets, max pooling, flattening, and dense layers. We adopt the design specifications in the literature \cite{bian2020respiratory} to create the ResNet architecture, including 1D convolution layers with ReLU activation. The input of the model is a PPG signal, a vector of shape (2048,1)-- which is a 32s segment with a sampling rate of 64. The model training is carried out with three different learning rates reported in \cite{bian2020respiratory}, and the best result is obtained for a learning rate of $10^{-5}$.

\textbf{CycleGAN}: A cycle-generative adversarial network was proposed by Aqajari et al. \cite{aqajari2021end} to reconstruct respiration signals from PPG, exploiting the reconstruction ability of the network. They first used an image translator to convert the PPG and respiration signals into images. Then, the model is trained with the images. Besides, to ensure that the reconstructed signal preserves the main features of the respiratory signals, they used adversarial losses, cycle consistency losses, and RR loss as part of the optimizing function. The architecture and specific network configurations include stride-2 convolutions, residual blocks, and fractionally-strided convolutions for the generative networks and 70$\times$70 PathGANs for the discriminator networks. In addition, the BreathMetrics library \cite{noto2018automated} was used to identify the peak values of the respiration signal and estimate the RR using those values.

\textbf{A-P Synthesis}: A hybrid method was introduced in \cite{huang2021novel} for estimating respiration rate (RR) from both PPG and ACC signals. The algorithm involved two independent steps for deriving RR estimates. First, a PPG-based RR algorithm was developed to extract surrogate respiratory signals, such as AM, FM, and BW, from the PPG signal and estimate RR from the reciprocal of the average of the most recent 30 inner breath intervals. Second, an ACC-based RR algorithm was introduced to project the three orthogonal ACC waveforms to the first principal axes and implement FFT spectrum and peak tracking to derive RR estimates. Final RR values were calculated by combining the PPG and ACC estimates considering the quality of the estimates. The final estimate would be discarded if both methods provided low-quality estimates. It is important to highlight that in this method, there are no machine learning algorithms involved, which means that the number of parameters is simply zero.

\subsection{Evaluation Measures}
To evaluate the accuracy of the predictions, we use the Mean Absolute Error (MAE), Root Mean Square Error (RMSE), and Bland Altman method \cite{dougan2018bland}. These metrics are employed to measure the discrepancy between the obtained RR and the ground truth RR. The MAE is computed by averaging the absolute differences between the values generated by the model and the ground truth. Moreover, the RMSE is obtained by taking the root square of the squared mean of residuals.
The MAE and RMSE are defined as follows:

\begin{equation}
MAE = \frac{1}{N}\sum_{i=1}^N|RR_e^i - RR_r^i|
\end{equation}
\begin{equation}
RMSE =\sqrt{ \frac{\sum_{i=1}^N|RR_e^i - RR_r^i|^2}{N}}
\end{equation}
where N refers to the total number of respiratory rates. $RR_e^i$ and $RR_r^i$  denote the calculated estimates and reference respiratory rates, respectively, for each segment.  The Bland Altman assesses the level of agreement between the estimated RR and the ground truth RR. This analysis provides a more comprehensive evaluation of the performance of the methods. 

\subsection{Test Set Results}
As previously mentioned, the performance of the methods is evaluated using the raw PPG, ACC, and GYR signals collected from 10 participants (i.e., test dataset). The estimates are compared with the ground truth RR collected from the chest-band device. Table~\ref{tab1} shows the summary of the proposed method performance compared with the state-of-the-art methods. As indicated, the proposed method outperforms the existing RR estimation methods in terms of MAE and RMSE, with the values of 1.85 $\pm$ 0.4 and 2.34 $\pm$ 0.3, respectively. The SF and A-P Synthesis methods obtained the highest MAE values, and the A-P Synthesis had the worst RMSE value.



In addition to MAE and RMSE metrics, we used Bland Altman analysis to assess the methods. As shown in Table~\ref{tab1}, the proposed method obtained the lowest mean bias (i.e., -0.63), indicating a superior level of agreement between its RR estimates and the reference RR when compared to the estimates from other methods. In the context of the Bland-Altman assessment method, the method with the smallest absolute mean bias (regardless of the sign) indicates a stronger agreement with the reference method, signifying a more accurate and reliable measurement. The CycleGAN obtained the highest mean bias. In contrast, the SF method obtained the most narrow confidence interval, showing the lowest level of uncertainty in RR estimation between these methods.
\begin{figure}[htbp]
\begin{center}
   \includegraphics[height=5.25cm ,width=9cm]{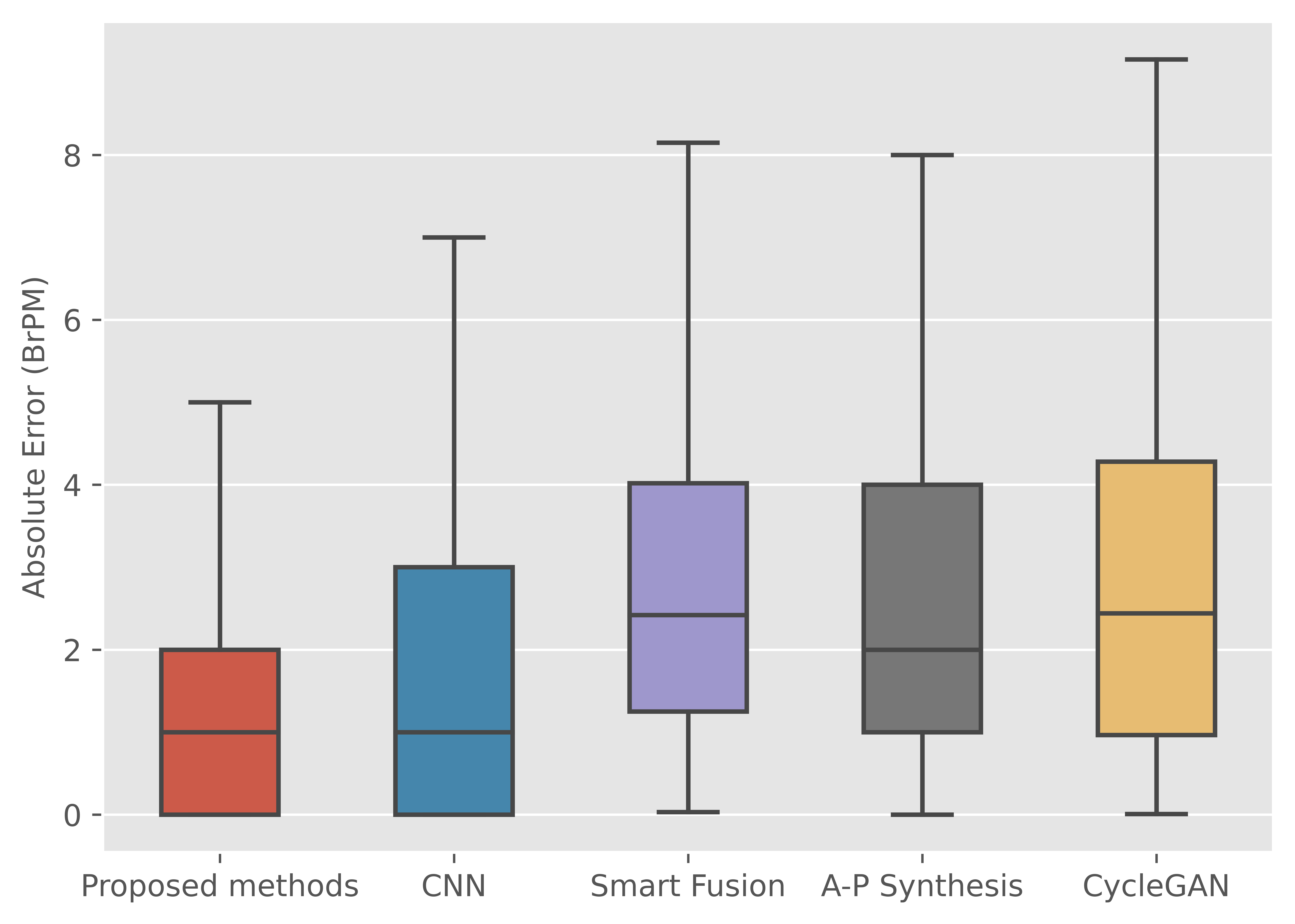}
    \caption{Box plot analysis of  Methods vs. Absolute Error.}
    \label{fig:4} 
\end{center}

\end{figure}

For a deeper evaluation, we also obtain the box plots of the methods, showing the distribution of the obtained absolute errors. The box plots are demonstrated in Fig.~\ref{fig:4}. Both the proposed method and the CNN method achieved a similar median absolute error of approximately 1.6 brpm (breathing rate per minute) for RR estimation. In contrast, the SF, CycleGAN, and A-P Synthesis methods had higher median absolute errors of over two brpm. It is noteworthy that the absolute error distribution of the proposed method is narrower (lower third quartile value) compared to the error distribution of the CNN method, indicating the effectiveness of our approach in providing more accurate RR estimation.

We also obtain the parameter counts, indicating the number of learned parameters in the models, which can be served as a measure of the method's complexity. The total parameter for the proposed method is 3,120,454, which is considerably lower than the parameter count of CycleGAN: i.e., 17,771,304. Compared to the proposed method, the CNN and SF methods had lower parameter counts as 457,037 and 1,900,234, respectively.

In summary, the proposed method shows superior performance compared to the other methods. Our findings clearly indicate the effectiveness of the proposed method in accurately estimating RR values from the signals collected from the PPG, ACC, and GYR of smartwatches. While the state-of-the-art methods may obtain a reasonable performance for RR estimation from fingertips-based PPG data collected under stationary conditions, their accuracy significantly decreases when used in real-life scenarios involving wrist-based PPG signals.

\section{CONCLUSION and Future work}\label{conclusion}
In this paper, we proposed a deep learning-based method for RR estimation using raw smartwatches PPG, ACC, and GYR signals. The proposed method consisted of three stages. The first stage included segmentation and filtering methods to extract 32-second raw PPG, ACC, and GYR segments and discard distorted signals due to user's physical activity. Second, respiratory-induced signals were extracted from ACC and GYR signals using ICA method and were scaled between -1 and 1. Third, a CNN composed of Multi-Scale Convolution augmented with dilated residual inception, 1D convolution layers, and dense layers was developed to estimate RR using PPG and the respiratory signals. The dilated residual inception module enhanced the efficiency of CNN by providing a large receptive field, and the Multi-Scale Convolution improved the Multi-Scale feature extraction ability and significantly reduced the vanishing gradient problem. We evaluated the proposed method using PPG, ACC, and GYR data collected via Samsung watches from 36 individuals for one day. We compared the proposed method with four existing RR estimation methods. Upon comparison with RR references obtained from a chest-band device, the proposed method obtained the best MAE and RMSE (i.e., 1.85 and 2.34), while the second best MAE and RMSE were 2.41 and 3.29 achieved by the CNN method. Additionally, our method obtained the smallest mean bias (i.e., -0.63) according to the Bland-Altman method, as well as a narrower absolute error distribution (with the smallest median) compared to the existing method. Consequently, our findings showed that our method outperformed the other methods and could estimate RR accurately. As future work, we will optimize the RR estimation method to reduce its computational requirements, enabling implementation on wearable and edge devices. We intend to decrease the model's dimensions while maintaining its precision by employing techniques, such as Knowledge Distillation.

\ifCLASSOPTIONcaptionsoff
  \newpage
\fi

\end{document}